\documentclass{JINST}
\usepackage{graphicx}
\usepackage{amssymb}
\usepackage{amsmath}
\usepackage{bm}

\title{Neutron calibration facility with an Am-Be source for pulse shape discrimination measurement of CsI(Tl) crystals}

\author{H.S.~Lee$^a$~\thanks{Corresponding author.}, H. Bhang$^b$, J.H.~Choi$^b$, S.~Choi$^b$, I.S.~Hahn$^c$, E.J.~Jeon$^d$, H.W.~Joo$^b$, W.G.~Kang$^d$, G.B.~Kim$^{b,d}$, H.J.~Kim$^e$, K.W.~Kim$^b$, S.C.~Kim$^b$, S.K.~Kim$^b$, Y.D.~Kim$^{d,f}$, Y.H.~Kim$^{d,g}$, J.H.~Lee$^b$, J.K.~Lee$^b$, D.S.~Leonard$^h$, J.~Li$^d$, S.S.~Myung$^b$, S.L.~Olsen$^b$, and J.H.~So$^d$ \\
\llap{$^a$} Department of Physics, Ewha Womans University, Seoul 120-750, Korea\\
\llap{$^b$} Department of Physics and Astronomy, Seoul National University, Seoul 151-747, Korea\\
\llap{$^c$} Department of Science Education, Ewha Womans University, Seoul 120-750, Korea\\
\llap{$^d$}Center for Underground Physics, Institute for Basic Science (IBS), Daejon 305-811, Korea\\
\llap{$^e$}Department of Physics, Kyungpook National University, Daegu 702-701, Korea\\
\llap{$^f$}Department of Physics, Sejong University, Seoul 143-747, Korea\\
\llap{$^g$}Korea Research Institute of Standards and Science, Daejon 205-340, Korea\\
\llap{$^h$}Department of Physics, The University of Seoul, Seoul 130-743, Korea\\
 E-mail: \email{hyunsulee@ewha.ac.kr}}

\abstract{
We constructed a neutron calibration facility based on a 300-mCi Am-Be source in conjunction with a
search for weakly interacting massive particle candidates for dark matter.  The facility is used to
study the response of CsI(Tl) crystals to nuclear recoils induced by neutrons from the Am-Be source
and comparing them with the response to electron recoils produced by Compton scattering of
662-keV $\gamma$-rays from a $^{137}$Cs source.  The measured results on pulse shape discrimination~(PSD)
between nuclear- and  electron-recoil events are quantified in terms of quality factors.  A
comparison with our previous result from a neutron generator demonstrate the feasibility of performing 
calibrations of PSD measurements using neutrons from a Am-Be source.
}


\begin{document}

\section{Introduction}
It is well known that the largest component of the gravitationally interacting matter in the Universe
is not ordinary matter but nonbaryonic exotic dark matter.  Weakly Interacting Massive Particles~(WIMPs)
are strong candidates for this dark matter~\cite{bwlee,Jungman}.  WIMPs are expected to interact with
ordinary matter via elastic scattering from nuclei and could be directly detected by measuring the
recoil energy of the nuclei that are expected to be in the \textless\ 100~keV~\cite{wimpdm} energy
range.  Because of the high stopping power of recoil nuclei with these energies, the light yield of
a scintillation detector is typically a factor of  5--10 times lower than that produced by 
$\gamma$-rays of equivalent energy. This reduction in light output is parameterized by a quantity
called the quenching factor. As a result of this, WIMP interactions in typical scintillation detectors
produce signals with visible energies of only a few keV.

Due to the effects discussed above, it is crucial for WIMP search experiments to have a quantitative
understanding of the detector response to low-energy recoiling target nuclei.
It is also important to be able to reduce the environmental backgrounds that are mostly due to $\gamma$ and
$\beta$ decays by exploiting differences in the detector responses to nuclear- and
orbital-electron-recoil-induced signals in the detector.
The high stopping power of a recoil nucleus produces a high density of ionization that results
in a time profile for photoelectron production that is different from that produced 
by electron recoils~\cite{birks}.  This characteristic property makes particle
identification possible using the time structure of the signal pulse.  The pulse shape
discrimination~(PSD) capabilities of various scintillating counters have been previously
reported~\cite{naipsd1,naipsd2,csipsd1,csipsd2,cdwopsd,pbwopsd,otherpsd}.

Among various scintillating crystals,  CsI(Tl) has special merits owing to its relatively large light
yields and good PSD capability~\cite{csipsd1,csitest}. Also,  large volume CsI(Tl) crystals are relatively
easy to fabricate, in part because they are not very hygroscopic.  The large atomic masses of
Cs and I nuclei enhance the cross section for coherent WIMP--nucleus spin-independent interactions
(which depend on $A^2$).  Also, Cs and I nuclei have relatively large spin factors for protons~($S_p$)
that enhance the cross section for WIMP--proton spin-dependent interactions (which depend on $S_p^2$).
Several studies aimed at measuring the PSD capability of CsI(Tl) crystals have been performed with neutron
beams generated by neutron generators~\cite{csipsd1,csipsd4} and a $^{252}$Cf neutron source~\cite{csipsd2}. 
These studies demonstrated
the good PSD capability of CsI(Tl) crystals in comparison with those of NaI(Tl) crystals. 

Previous studies of the PSD capability of CsI(Tl) crystals in our group used mono-energetic neutrons produced by the
$^3$H(p,n)$^3$He reaction of a neutron generator at the Korea Institute of Geoscience and Mineral resources~(KIGAM)~\cite{csipsd3}.  
Even though this mono-energetic neutron beam facilitates the understanding of the deposited
energy in the crystal so that one can measure quenching factors, the high costs of construction and maintenance make them  impractical
for a university-based research facility. Our opportunities to use neutron beam generators at remote sites
were limited by available beam time and funding. 

For WIMP searches, however, we need to understand the dependence
of the PSD capability for each detector on many variables, 
such as the raw powder used for growing
the crystal, the growth method, the  company that grew the crystal, the photon sensor, and the electronic
readout, including the trigger and data-acquisition system.  
In order to measure the PSD power 
of the CsI(Tl) crystals used for our searches within time and cost constraints, we constructed a neutron
calibration facility based on a recycled 300-mCi Am-Be neutron source.  
There was another study that used
a neutron source~($^{252}$Cf) for the PSD measurement of the CsI crystal~\cite{csipsd2}. 
That measurement demonstrated the feasibility of the use of a neutron source for the PSD measurement. 
Our facility has a special design that includes $\gamma$ and neutron tagging detectors that 
can reduce possible contamination of the electron recoil events caused by environmental background as well
as accompanying $\gamma$s of the neutron source in neutron calibration data. 
We measured the
responses to nuclear recoils induced by neutrons from this source and compared them with the 
responses to electron recoils produced by a $^{137}$Cs $\gamma$ source.  The performance of this Am-Be neutron
calibration facility  for the PSD measurement was found to be compatible with that using  the neutron generator.

\section{Experiments}
\begin{figure}
\centering
\includegraphics[width=0.7\columnwidth]{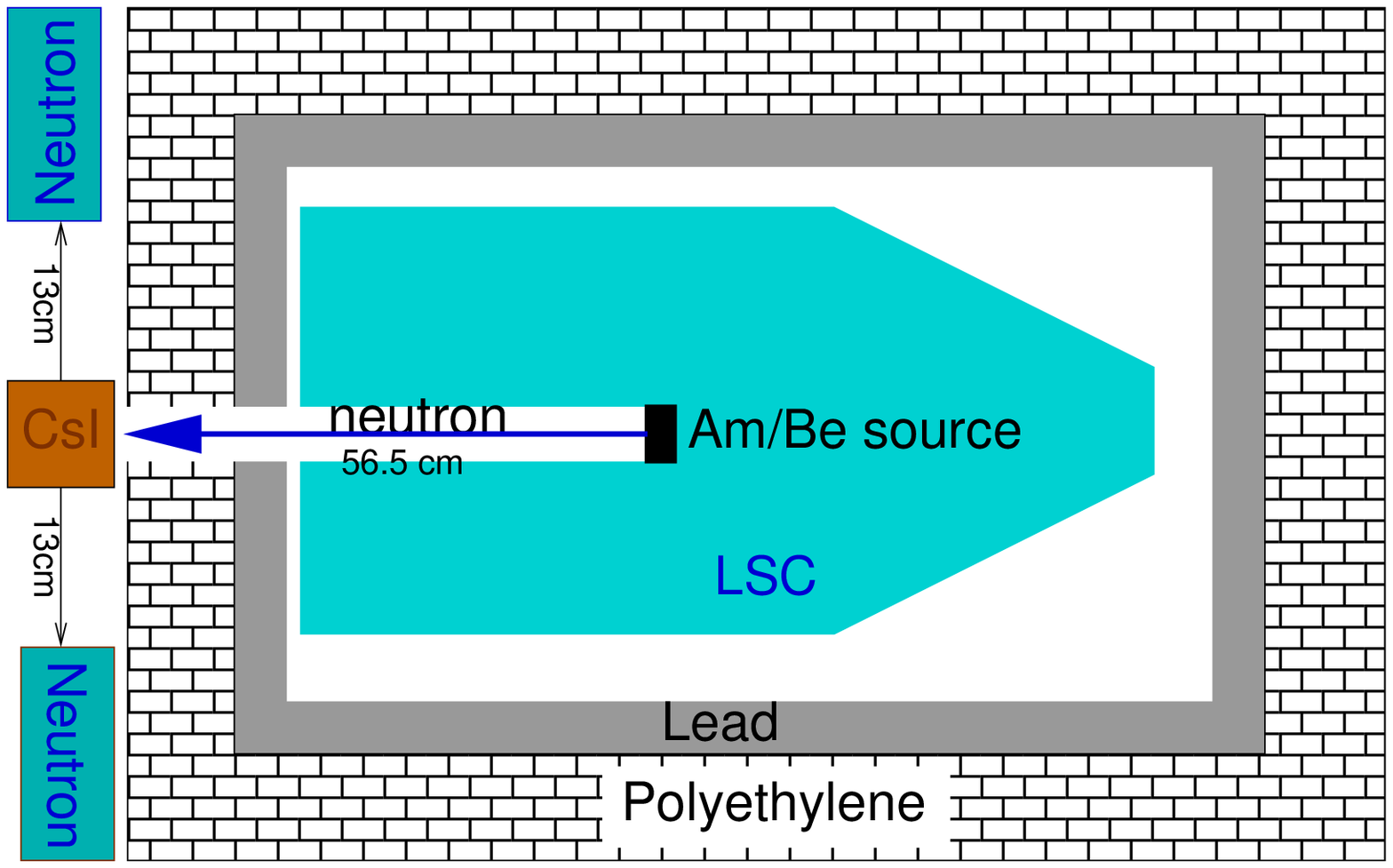}
 \caption{Neutron calibration setup.
} \label{fig:ana_nucal_setup}
\end{figure}

We constructed a neutron calibration facility at Seoul National University using a 300-mCi Am-Be source
recycled from Korea Cancer Center Hospital.  Alpha particles emitted by $^{241}$Am impinge on a $^9$Be
target and produce the ($\alpha$,$n$) reactions:
\begin{eqnarray*}
        \alpha + {^9\rm{Be}} & \rightarrow & ^{12}\rm{C} + n ~(\sim\negthickspace50\%), \nonumber\\
        \alpha + {^9\rm{Be}} & \rightarrow & ^{12}\rm{C}^*  + n  ~(\sim\negthickspace50\%), \nonumber\\
        &             & ^{12}\rm{C}^* \rightarrow {^{12}\rm{C}} +\gamma ~(4.4~\rm{MeV}).
        \label{eq:reaction}
\end{eqnarray*}
These neutrons were used to irradiate CsI(Tl) crystals with the arrangement depicted in
Fig.~\ref{fig:ana_nucal_setup}.  The neutron source is surrounded by liquid scintillator~(LSC)
composed of 95\% mineral oil and 5\% pseudocumene with a collimated aperture in the CsI(Tl) crystal
direction. The LSC acts as a tagging detector for the 4.4-MeV $\gamma$ particles as well as a passive
neutron shield.  Additional passive shielding for reducing uncollimated neutron and $\gamma$ fluxes 
to a level sufficient for safe operation, is provided by a structure comprised of 5~cm of lead and
10~cm of polyethylene.   The CsI(Tl) crystal under study is located at the exit of the collimated aperture
as shown in the figure.  To identify neutrons scattered from CsI, two neutron
tagging detectors are located at 90$^{\circ}$ from the line between the neutron source and the CsI(Tl)
crystal. The neutron detectors  consist of BC501A liquid scintillator contained in cylindrical 0.5~$\ell$
stainless-steel containers. Each neutron detector is read-out by a single, 2-inch photomultiplier tube
(PMT)~(H1161, Hamamatsu Photonics) and surrounded by 10-cm paraffin
and 5-cm lead blocks to reduce external background.

We used two different CsI(Tl) crystals from two different crystal-growing companies: a  $3\times 3\times 3$~cm$^3$
crystal from Beijing Hamamatsu~(Beijing crystal) and a $3\times 3\times 2$~cm$^{3}$ crystal from SCICCA in
Shanghai~(Shanghai crystal).  These two crystal companies grew full-sized~(8$\times$8$\times$30~cm$^3$)
CsI(Tl) crystals that are used for WIMP search experiments performed at the Yangyang Underground
Laboratory~\cite{kims1,kims2,kims3}.  Each of the small-sized crystal was grown in the same ingot with one of the
full-sized crystals.  Two 3-inch PMTs (D726UK, Electron Tubes) with green-extended RbCs photocathodes were
attached to each end of the CsI(Tl) crystal.  The other four crystal surfaces were covered by
two layers of Teflon~(0.2~mm thick) and black vinyl sheets.  The amplified signals from the crystal and
the neutron detectors were encoded by 400-MHz flash analog-to-digital converters~(FADCs) for a 32-$\mu$s
time interval.   The CsI(Tl) crystals' photoelectron~(PE) yields were measured to be $\approx$5~PE/keV  
using 59.54-keV $\gamma$-rays from an $^{241}$Am calibration source. The trigger condition for the CsI(Tl)
crystal was two or more PEs in each PMT within a 2-$\mu$s time window, corresponding to a total of four
or more  PEs in the detector, which corresponds to a visible energy threshold of $\sim$1~keV. The event rate
of the CsI(Tl) crystal during exposure to the neutron source was $\sim$300~Hz, mostly due to
background $\gamma$-rays produced in the source.  This trigger rate was reduced to  $\approx$0.3~Hz
by requiring  a time coincidence between the CsI(Tl) crystal and one of the neutron detectors within a
2~$\mu$s time window.

For the comparison with the electron recoil signals, the crystal was irradiated with 662-keV $\gamma$-rays
from a $^{137}$Cs source that induced Compton scattering.   These data were taken separately with the same setup
except for a shield between the crystal and the Am-Be neutron source. The CsI(Tl) trigger condition was the
same as that used for the neutron measurement, except for the time coincidence with one of
the neutron detectors.

\section{Analysis}
\subsection{Neutron identification}
\begin{figure}[!htb]
\begin{center}
\begin{tabular}{cc}
\includegraphics[width=0.48\textwidth]{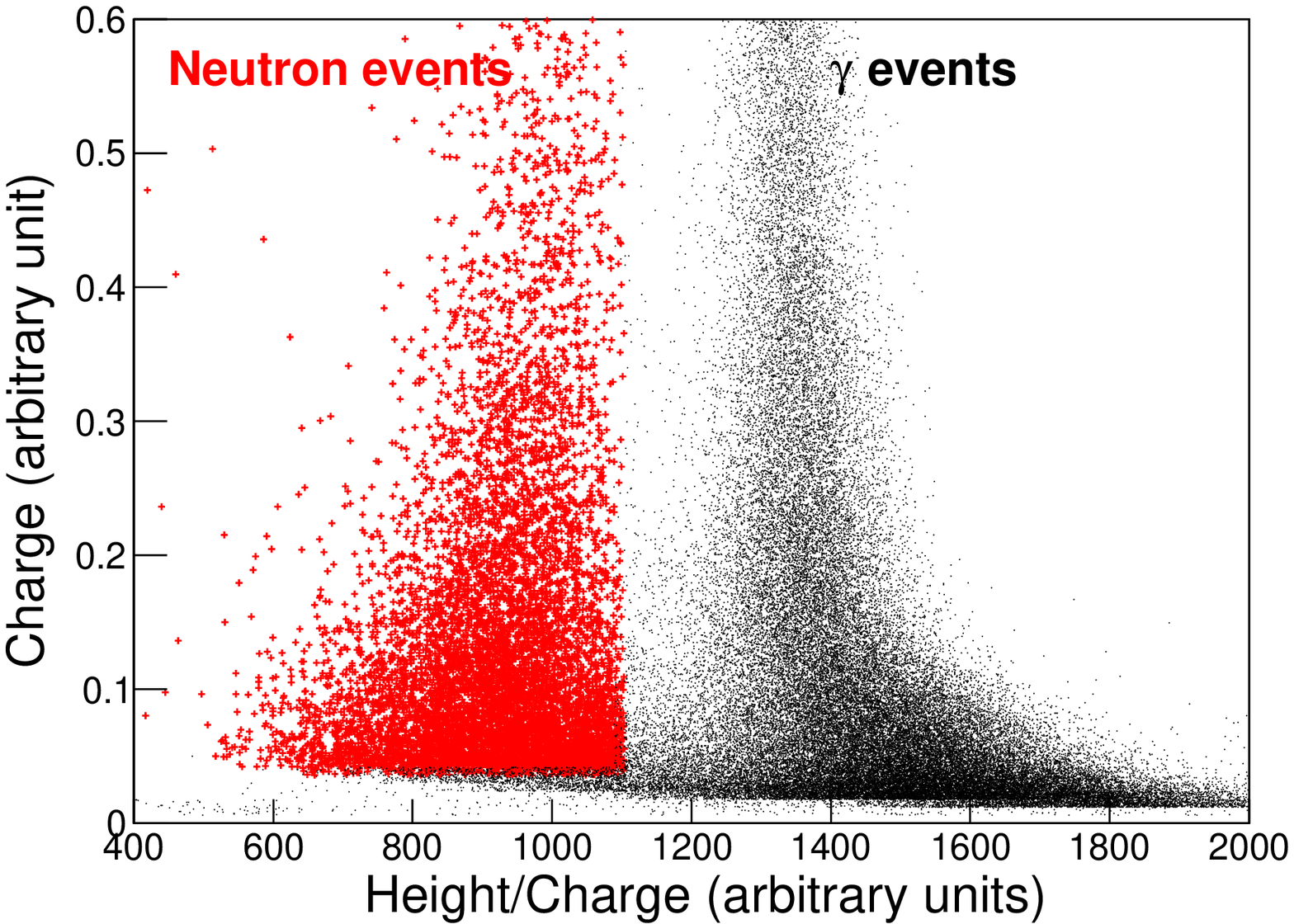}&
\includegraphics[width=0.48\textwidth]{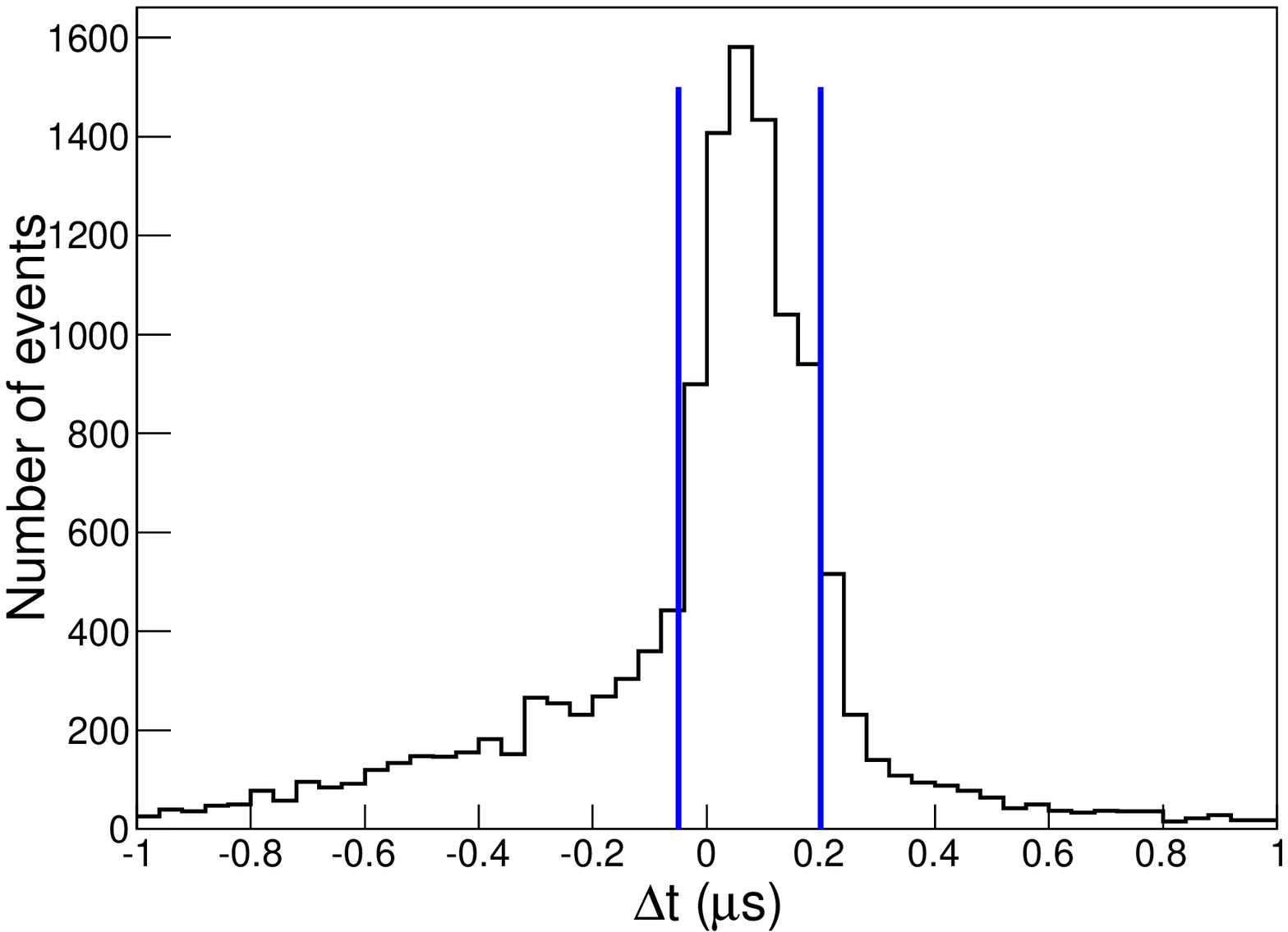}\\
 (a)   & (b)   \\
\end{tabular}
\end{center}
\caption{ (a) The ratio of the maximun pulse height to the total charge (horizontal) {\it versus}
the total charge for neutron-induced (solid red circles) and $\gamma$-induced (black points) 
for one of the neutron detectors. (b) Time difference between
the CsI(Tl) crystal start-time and that of one of the neutron detectors.
  }
\label{fig:nselection}
\end{figure}

It is well known that BC501A liquid scintillator can separate neutrons from $\gamma$ rays using
PSD~\cite{hjkim04,jjzhu} because neutron-induced signals have a longer tail than $\gamma$-induced events
with the same visible energy.  Therefore, by using the ratio of  total charge to the maximum pulse-height
of the signal,  neutrons are clearly separated from $\gamma$ rays, as can be seen in Fig.~\ref{fig:nselection}(a).

\begin{figure}[!htb]
\begin{center}
\includegraphics[width=0.7\textwidth]{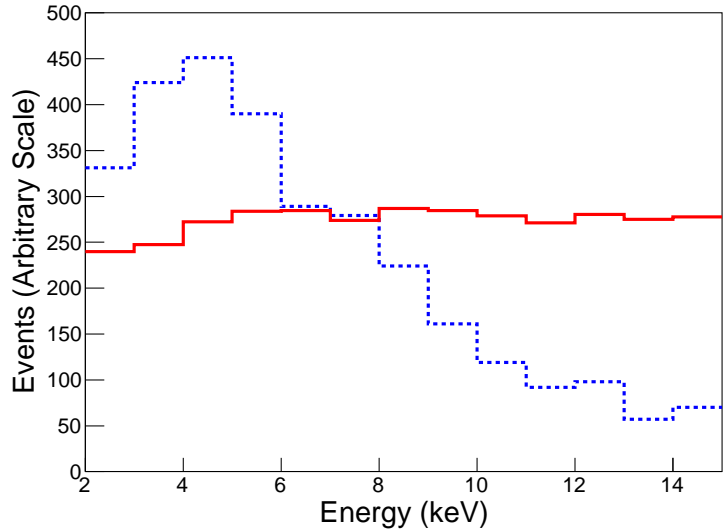}
\end{center}
\caption{The low energy distribution of the neutron-induced events from the Am-Be neutron source~(dashed line) and
the electron-induced events from the $^{137}$Cs source~(solid line) are shown. The spectrum of electron-induced events
is scaled down by a factor of 10.  
  }
\label{fig:edist}
\end{figure}

The 2~$\mu$s coincidence between the CsI(Tl) crystal and one of the neutron detectors was contaminated by
accidentals.  This background was reduced by a stricter  coincidence condition requirement that was set
off-line, as shown in Fig.~\ref{fig:nselection}(b).  Here $\Delta t$ is the the~CsI crystal start-time minus
the start-time~of~the neutron detector.  With good neutron separation capabilities and tighter time coincidence
of the neutron detectors, we recorded neutron calibration data with the Am-Be neutron source over a
two-month period for each crystal. Figure~\ref{fig:edist} shows the energy spectra of the neutron-induced 
events with the Am-Be source as well as the electron-induced events from $^{137}$Cs source. 

\subsection{CsI(Tl) data analysis}
Because the decay time of the scintillation light in the CsI(Tl) crystal is rather long, pulses from
single photoelectrons are well separated in low visible-energy events.  Single photoelectrons~(SPEs)
in an event were identified by the application of a clustering algorithm to the FADC data.  From
consecutive above-pedestal FADC bins, local maximum points are found and these are used
to identify isolated clusters. In the case in which two local maxima are found in the same
cluster of FADC bins,  neighboring clusters are separated by locating the local minimum point between
two clusters.  When this technique is applied to $^{55}$Fe source data, we find that in 5.9-keV signal
events \textgreater90\% of the SPEs are well identified as isolated single clusters while the remaining
SPEs are in overlapping clusters.  A detailed explanation and an evaluation of the performance of the
clustering algorithm can be found in Ref.~\cite{kims1}.  

\begin{figure}
\centering
\includegraphics[width=0.7\columnwidth]{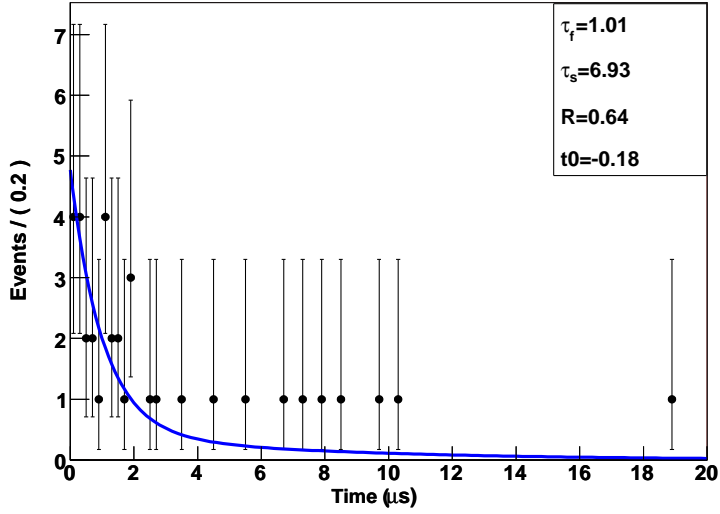}
\caption{
An example of the unbinned maximum likelihood fit of the time distribution of the identificed clusters
using the two exponential decay components. This event
corresponds to a neutron-induced event with a visible energy of $\sim$6-keV.
} \label{ana_decay_fit}
\end{figure}

The sum of the single-cluster charges over the
entire time window is used to determine the deposited energy.  Low-energy calibrations are done using
the $^{241}$Am  59.54-keV $\gamma$ peak. 
To characterize the pulse shapes of the nuclear- and
electron-recoils, an unbinned maximum likelihood fit for the time distribution of photoelectrons in an event
is performed with a double exponential
function of the form:
\begin{equation}
        f(t)=\frac{1}{\tau_{f}}e^{-(t-t_{0})/{\tau_{f}}}+\frac{R}{\tau_{s}}e^{-(t-t_{0})/{\tau_{s}}},
        \label{eq:ana_decay}
\end{equation}
where $\tau_{f}$ and $\tau_{s}$ are decay time constants for the fast and slow components, respectively, $R$ is
the ratio between the two components, and $t_{0}$ is the start time corresponding to the time of the first
cluster in an event. Figure~\ref{ana_decay_fit} shows an example of the fit result for a typical photoelectron
time distribution of an event. We performed an event-by-event fit for both the neutron source data and
the $^{137}$Cs source data. From the fitted parameters, we calculated the mean time~(MT) of each event as
\begin{equation}
\mathrm{MT}=\int t\cdot f(t)dt/\int f(t)dt,
\end{equation}
and used the MT as a PSD parameter to discriminate between nuclear-  and electron-recoil events.

\subsection{Pulse shape discrimination~}
It is possible to use PSD with the CsI(Tl) crystals because of the different time distributions of photoelectrons
between nuclear- and electron-recoils~\cite{knoll}.  We characterize the PSD power by a quality factor~\cite{gaiskell}
defined as
\begin{equation}
K \equiv \frac{\beta(1-\beta)}{(\alpha-\beta)^2},
\label{eq:quality}
\end{equation}
where $\alpha$ and $\beta$ are fractions of the nuclear- and electron-recoil events that
satisfy the selection criteria, respectively,  where we have varied these criteria to provide the
best~({\it i.e.}, the minimum) value of the quality factor.  For a detector with perfect discrimination,
$\alpha=1$ and $\beta=0$, thus a smaller quality factor means a better PSD power.

\begin{figure}
\centering
\includegraphics[width=0.7\columnwidth]{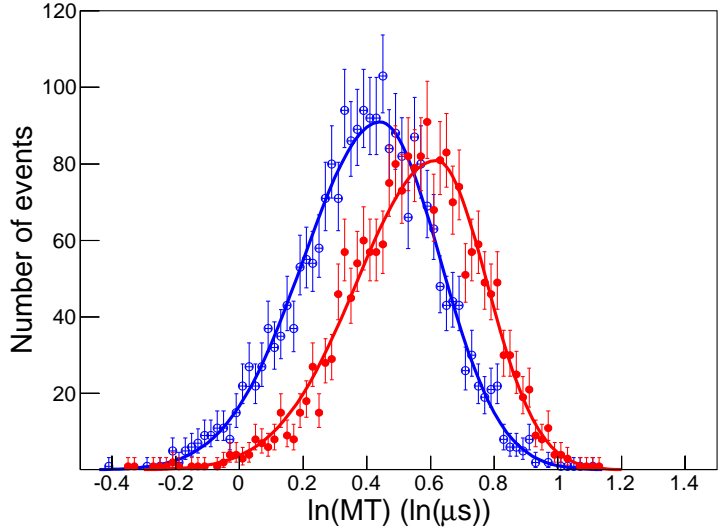}
\caption{The ln(MT) distribution for the Beijing crystal's response to neutron-induced events~(open circles) from the Am-Be source and
electron-induced events~(filled circles) from the $^{137}$Cs source with visible energies between 5 and 6~keV. Superimposed
are the results of a fit that uses asymmetric Gaussian functions.
} \label{fig:logmtex}
\end{figure}

For a simple separation parameter, we choose the natural logarithm of MT~(ln(MT)).  Figure~\ref{fig:logmtex}
shows the ln(MT) distributions for nuclear and electron recoil events in the Beijing
crystal with visible energies in the range 5~$<E<$6~keV.  It is evident in this figure that the decay time for nuclear
recoils is shorter than that for electron recoils, consistent with previous measurements~\cite{csipsd1,csipsd2,csipsd4,csipsd3}.  We performed fits to the ln(MT) distributions
using asymmetric Gaussian functions with $x=\ln(\text{MT})$ as follows:
\begin{eqnarray*}
g(x) = &
\frac{1}{1/2(\sigma_L+\sigma_R)}e^{-\frac{1}{2}\left(\frac{x-m}{\sigma_L}\right)^2}, &x<m, \\
&\frac{1}{1/2(\sigma_L+\sigma_R)}e^{-\frac{1}{2}\left(\frac{x-m}{\sigma_R}\right)^2},&x \geq m,\\
\end{eqnarray*}
where $\sigma_L$ and $\sigma_R$ are standard deviations of left- and right-side Gaussians, respectively, and
$m$ is the most probable value.  This parameterization is slightly different from that in previous reports in which
a simple Gaussian distribution was assumed~\cite{csipsd2,csipsd3}.  However, in this analysis we employed a
different method to evaluate ln(MT) and used a much longer time window.  Our data were well described
by asymmetric Gaussians function, as can be seen in Fig.~\ref{fig:logmtex}.

\begin{figure}[!htb]
\begin{center}
\begin{tabular}{cc}
\includegraphics[width=0.48\textwidth]{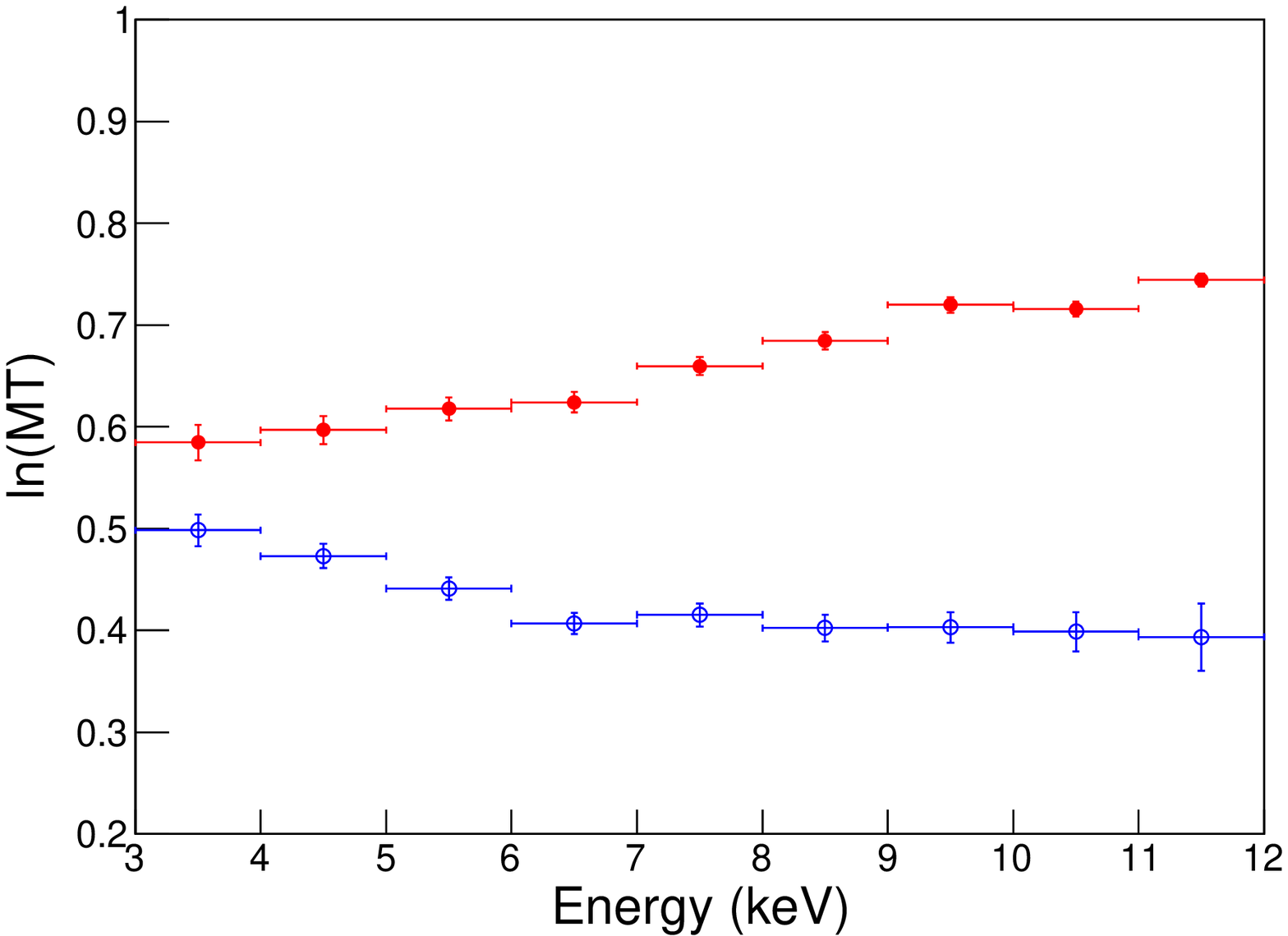}&
\includegraphics[width=0.48\textwidth]{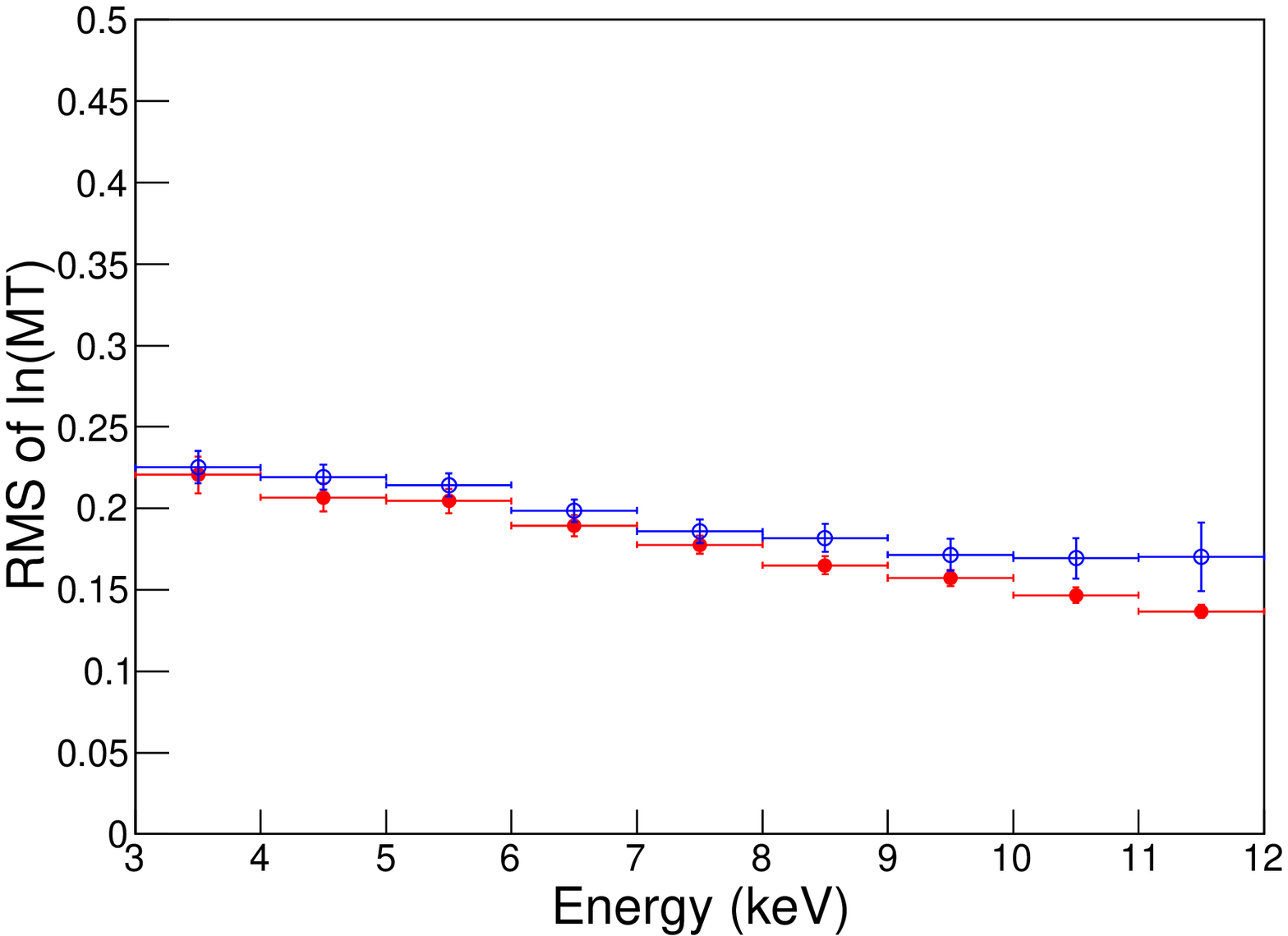}\\
(a) & (b) \\
\end{tabular}
\end{center}
\caption{(a) The most probable value of ln(MT) from the asymmetric Gaussian fits for each energy bin for neutron-induced
(open circles) and electron-induced (filled circles) events from the Am-Be source and the $^{137}$Cs source, respectively, 
for the Beijing crystal. (b) The root mean square~(RMS)
of ln(MT) as a function of energy for neutron-induced (open circles) and electron-induced (filled circle) events. 
} \label{fig:logmt}
\end{figure}

We performed the fit for each 1-keV energy bin from 3 to 12~keV for both neutron and $\gamma$ calibration
data.  Figure~\ref{fig:logmt} shows the most probable value and the root mean square~(RMS) of ln(MT) from the asymmetric Gaussian
fit of two different sets of calibration data.  There is an energy dependence to the most probable values as well as the RMS values: at low
energy, the separation power is reduced but we still retain  some PSD capability at energies as low as 3~keV, a
result that is similar to our previous measurement using  the neutron generator~\cite{csipsd3}.

\begin{figure}
\centering
\includegraphics[width=0.7\columnwidth]{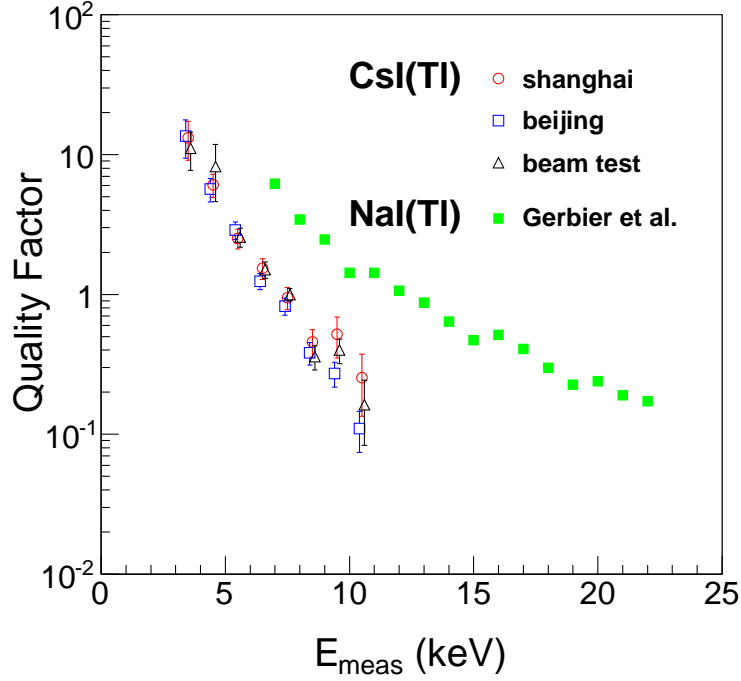}
\caption{The extracted quality factors of the CsI(Tl) crystals  compared to  neutron generator data~\cite{csipsd3} and  NaI(Tl) measurements~\cite{naipsd1}.
} \label{fig:quality}
\end{figure}

To calculate the qaulity factors in Eq.~\ref{eq:quality}, we provide simple selection criteria for ln(MT) as ln(MT)$<c$.
We have varied these criteria $c$ in range between -0.5 and 1.5 to find the minimum value of the quality factor that maximizes the nuclear- to electron-recoil ratio in each keV bin events.
Figure~\ref{fig:quality} shows the extracted (minimum) quality factors for the two CsI(Tl) crystals with the Am-Be
source compared with our previous result using the neutron generator as described in Ref.~\cite{csipsd3},  where it is apparent
that the quality factors obtained using the neutron source are consistent with those found using the
neutron generator.  This  demonstrates that we can use the
Am-Be neutron source for the PSD measurement of the CsI(Tl) crystal.  
We also  find that the quality factors of the CsI(Tl) crystals are  a factor
of $\sim$10 lower than those of  NaI(Tl) crystals.

\section{Conclusions}
We constructed a neutron calibration facility using the recycled 300-mCi Am-Be source to study the response
of CsI(Tl) crystals to nuclear recoil events. Even though this facility can not measure the quenching factor,
we measure the PSD parameters between the nuclear
recoil events and the electron recoil events of two CsI(Tl) crystals.  The extracted quality factors are
consistent with our previous measurements using the neutron generator.  This gives us
confidence in the use of this neutron calibration facility for the PSD measurements.
The neutron calibration data of the two CsI(Tl) crystals obtained with this neutron calibration facility
have already  been used to extract the nuclear recoil events of WIMP search data obtained at the Yangyang
Underground Laboratory~\cite{kims2,kims3,kimslow}.  This neutron calibration facility is also being used
to evaluate PSD capabilities of other candidate WIMP search scintillation detectors.

\acknowledgments
This research was funded by Grant No. IBS-R016-D1 and was supported by the Basic Science Research Program through the National Research Foundation of Korea (NRF) funded by the Ministry of Education (Grant No. NRF-2011-35B-C00007).

\end{document}